# Topological Skyrmion-type microparticle manipulation based on surface acoustic wave phase modulations

**Jiaqi Zhang[#], Decai Wu[#], Tingfeng Ma[*], Chenbowen Lou, Bowei Wu, Shuanghuizhi Li**

Zhejiang-Italy Joint Lab for Smart Materials and Advanced Structures, School of Mechanics and Engineering Science, Ningbo University, Ningbo 315211, China;

## Abstract

Surface acoustic wave (SAW) micromanipulation enables the precise, non-contact handling of microscale particles and has attracted considerable interest in microfluidics and biomedicine. However, conventional SAW platforms generally rely on simple interference fields which are susceptible to fabrication imperfections and environmental perturbations, resulting in limited trapping stability. Here, we develop a SAW-based acoustofluidic platform that generates an acoustic skyrmion lattice through the coherent interference of three SAWs. The topologically structured field provides robust phase singularities and a stable gradient-force landscape, enabling microparticles to be localized at predefined lattice sites and supporting controllable rotational manipulation. Independent modulation of the amplitude and phase of the electrical inputs allows the field strength to be tuned for particles of different sizes. Numerical simulations and proof-of-concept experiments confirm particle trapping and ordered lattice assembly in the acoustic skyrmion field, demonstrating the feasibility of translating topological acoustic textures into practical on-chip manipulation functions. This reconfigurable strategy offers a route to robust SAW manipulation and may support applications in single-cell analysis, three-dimensional cell assembly, high-throughput screening, and microscale and nanoscale device assembly.

[#] These authors contributed equally to this work.

[*] Authors to whom correspondence should be addressed: matingfeng@nbu.edu.cn

## 1. Introduction

Surface acoustic wave (SAW)-based particle manipulation is a central technology in acoustofluidics. SAWs are typically generated by interdigital transducers (IDTs) patterned on a piezoelectric substrate and couple to fluids and suspended particles through acoustic radiation forces and acoustic streaming.[1,2] These interactions enable precise, contact-free manipulation of microscale and nanoscale objects.[2–4] Recent studies and reviews have highlighted the versatility of acoustic manipulation for particle trapping, patterning, separation, and transport.[5–9] Representative applications include three-dimensional single-cell manipulation, high-throughput cell sorting, and acoustic virus isolation.[10–13] Most current SAW platforms, however, rely on conventional standing waves or simple interference patterns. Although electronic phase control can translate pressure nodes and particle streams with high precision,[14,15] these trapping landscapes remain dependent on ordinary interference conditions and can be sensitive to phase errors, coupling variations, and local perturbations. This motivates the development of robust structured acoustic fields for complex operating conditions.

Topological acoustic structures offer a potential route to overcoming these limitations. In particular, acoustic skyrmions combine robustness against local defects with distinctive spin-orbit-coupled wavefield textures, making them promising for stable particle trapping and controllable rotation.[16,17] Topologically structured wavefields recently begun to be explored for particle manipulation. For example, Wang *et al.* constructed topological wave structures, including skyrmions, in a gravity-wave system and demonstrated the trapping, orbital motion, and self-rotation of floating particles.[18] This study established the feasibility of using topological wavefields to control particle motion. Nevertheless, the application of acoustic skyrmions to SAW-based particle manipulation remains largely unexplored. In particular, the enhanced stability and robustness expected from their non-trivial topology have not yet been systematically validated in SAW systems, while suitable device architectures and the underlying particle-manipulation mechanisms require further investigations.

Motivated by these advances, we introduce the concept of acoustic skyrmions into a SAW platform and propose a topologically structured particle manipulation method based on three-wave interference. A full simulation of a SAW device comprising a piezoelectric substrate and a liquid-filled microchannel would require computationally intensive multiphysics coupling. This challenge is compounded by the short wavelength of SAWs, which substantially increases the required spatial resolution and computational cost. Because the fundamental principles governing coherent-wave interference are generally applicable to mechanical waves, we employ a reduced theoretical model to investigate the formation and modulation of the acoustic skyrmion field. The analysis therefore focuses on the physical mechanism underlying the generation and control of the topologically structured wavefield.

In this work, we designed an IDT device on a Z-cut $LiNbO_3$ piezoelectric substrate. Selective excitation of the corresponding electrode sets generates three mutually coherent SAWs whose interference produces an acoustic skyrmion lattice. The robust field gradients and local acoustic angular momentum associated with this structured field enable microparticles to be trapped at predefined lattice sites and provide a mechanism for orbital motion and controllable self-rotation. We first establish the theoretical basis of standing-wave fields and multiwave interference and then describe the design principles of the topological acoustic tweezer device. Finally, numerical simulations and experiments are used to evaluate the particle manipulation performance of the acoustic skyrmion field. This work provides an experimental basis and a practical device strategy for exploiting acoustic skyrmions in SAW-based particle manipulations. It also offers a potential route towards robust microparticle control and may facilitate future applications in precise cell manipulation, biomedical analysis, and micro- and nanoscale device assembly.

## 2. Theory and simulation of acoustic-field construction based on phase modulations

Consider a regular dodecagonal domain, as shown in Fig. 1. Each boundary can emit a plane wave directed toward the center, with identical amplitude and frequency and an independently adjustable phase.

When two or more acoustic waves of the same frequency but different propagation directions overlap in space, they form an acoustic interference field. For two counter-propagating waves along the $x$ direction, the pressure distribution of the resulting standing wave exhibits periodic antinodes and nodes. The acoustic pressure can be written as:

$$p(x,t) = p_0 \cos(kx - \omega t) + p_0 \cos(-kx - \omega t) = 2p_0 \cos(kx) \cos(\omega t), \qquad (1)$$

Setting $p(x,t) = 0$, the acoustic-pressure nodes are located at $x = (2n+1)\lambda/4$; similarly, setting $p(x,t) = 2p_0$, the antinodes are located at $x = n\lambda/2$. Under the acoustic radiation force, the microparticles form a one-dimensional periodic array.

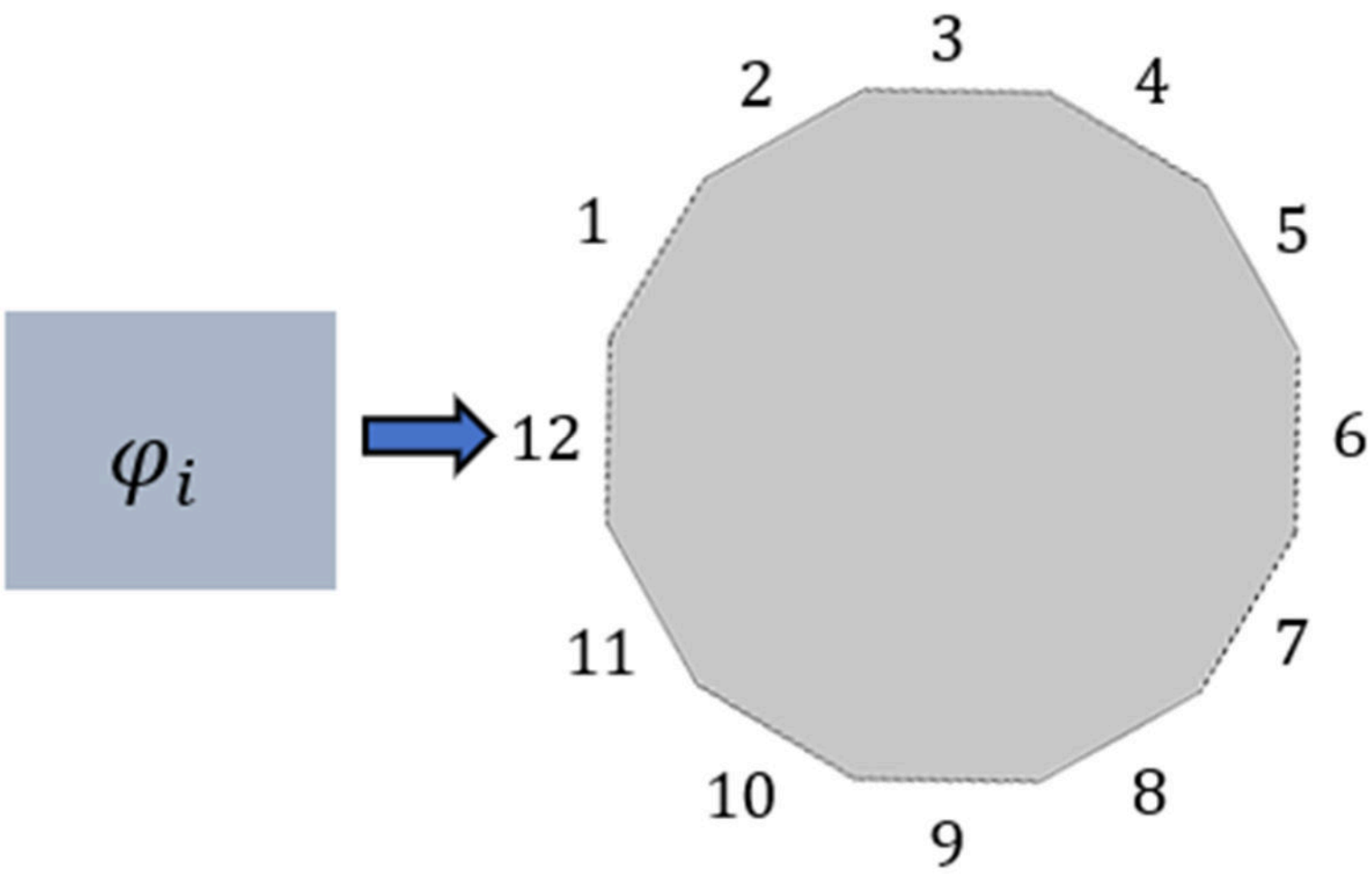

Fig. 1. Phase-tunable acoustic field

When the phases of the two incident waves are included, the acoustic-pressure field is written as:

$$p(x,t) = p_0 \cos(kx - \omega t + \varphi_1) + p_0 \cos(-kx - \omega t + \varphi_2), \tag{2}$$

Using the trigonometric identity $\cos \mathrm{A} + \cos \mathrm{B} = 2\cos\frac{A-B}{2}\cos\frac{A+B}{2}$, substitution gives the standing-wave equation:

$$p(x,t) = 2p_0 \cos(kx - \tfrac{\varphi_2-\varphi_1}{2}) \cos(kx + \tfrac{\varphi_2+\varphi_1}{2}), \tag{3}$$

The standing-wave envelope determines the positions of the antinodes and nodes:

$$|p(x)| = |2p_0 \cos\left(kx - \tfrac{\varphi_2-\varphi_1}{2}\right)|, \tag{4}$$

Setting $|p(x)| = 0$ gives the nodal condition. Solving for the node positions yields:

$$x_n = \frac{\varphi_2-\varphi_1}{2k} + \frac{(2n+1)\pi}{2k}, n = 0, \pm 1, \pm 2, \dots \dots, \tag{5}$$

Equ. (5) shows that standing-wave nodes translate with the phase difference. Accordingly, continuous and precise control of the nodal positions can be achieved by varying the phase difference between the incident waves. As shown in Fig. 2, two waves of the same frequency and adjustable phase interfere to form a one-dimensional standing-wave field. As their initial phase difference varies from 0 to $2\pi$, the white dashed line tracks the moving node. Increasing the phase difference from 0 to $2\pi$ moves the node continuously upward, whereas varying it from 0 to $-2\pi$ moves the node continuously downward.[14,15]

In addition, an instantaneous phase-difference switch from $\Delta\varphi = 0$ to $\Delta\varphi = \pi$ causes microparticles at the original standing-wave nodes to converge or separate.

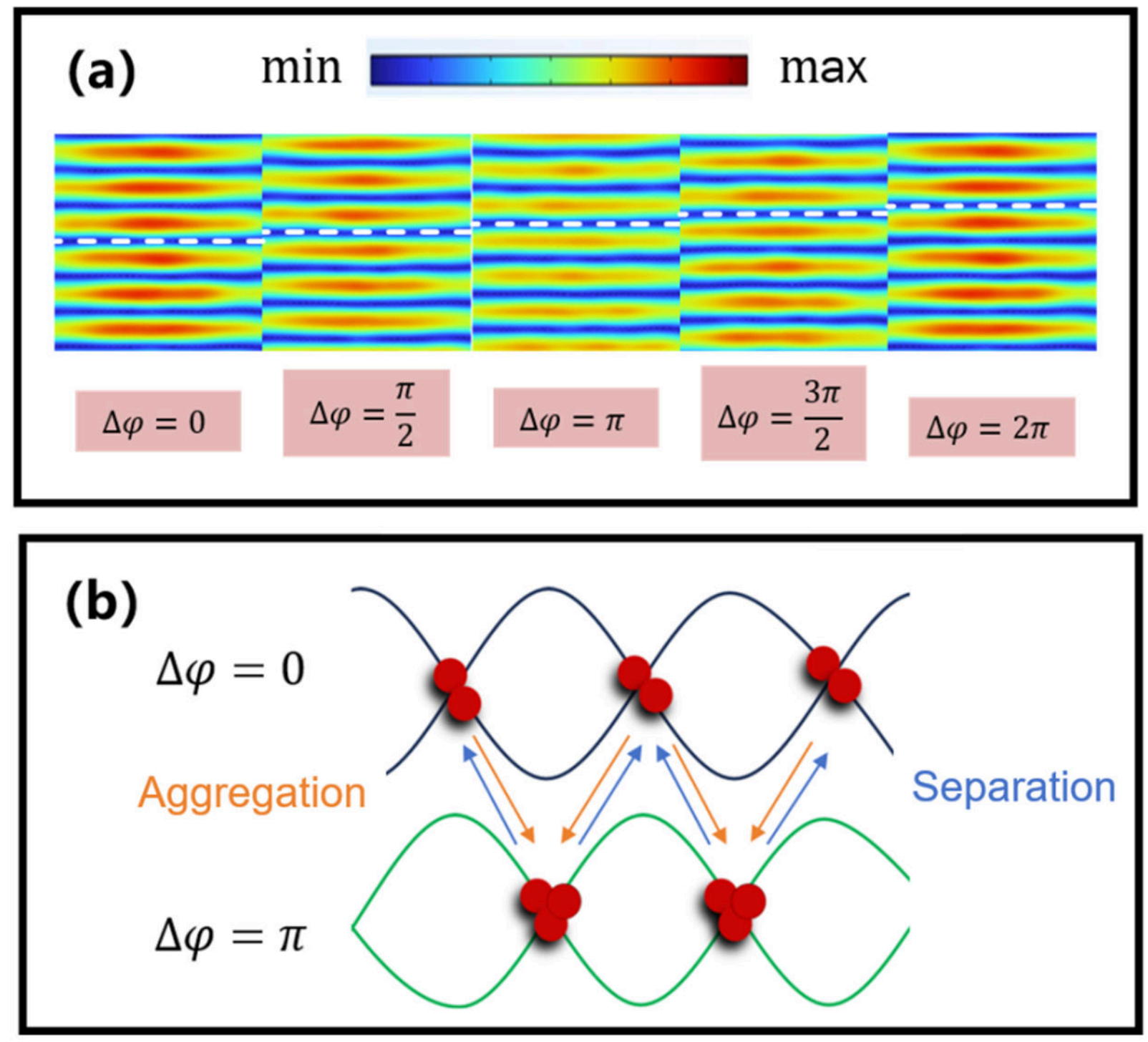

Fig. 2. Variation of one-dimensional standing-wave nodes with phase: (a) continuous phase-difference variation produces continuous translation; (b) an instantaneous phase-difference switch induces microparticle convergence and separation.

When acoustic waves propagate in multiple directions, the interference pattern becomes more complex. For N equal-amplitude, equal-frequency plane waves with wave vectors $k_i$, the total acoustic-pressure field can be expressed as:

$$p_i(\mathbf{r}, \mathrm{t}) = p_0 e^{i(\boldsymbol{k_i} \cdot \boldsymbol{r} - \omega t + \varphi_i)}, \tag{6}$$

where $p_0$ is the incident-wave amplitude, $\boldsymbol{k_i}$ is the wave vector of the ith wave, and $\varphi_i$ is its initial phase. The total field is the coherent superposition of all incident waves:

$$p(\mathbf{r}, \mathrm{t}) = \sum_{i=1}^{N} p_i(\mathbf{r}, \mathrm{t}) = p_0 e^{-i\omega t} \sum_{i=1}^{N} e^{i(\boldsymbol{k_i} \cdot \boldsymbol{r} + \varphi_i)}, \tag{7}$$

Taking the real part gives the acoustic pressure:

$$p_{total} = \mathrm{Re}\left[p_0 \sum_{i=1}^{N} e^{i(\boldsymbol{k_i} \cdot \boldsymbol{r} - \omega t + \varphi_i)}\right], \tag{8}$$

The intensity distribution of the interference field is determined by the phase differences among the constituent waves. By designing the wave-vector directions and phase distribution, an acoustic field with a prescribed topological structure can be constructed.

Skyrmion is a topologically stable vector-field structure whose unit vector covers the entire unit sphere.[16] In an acoustic system, an acoustic skyrmion can be constructed through three-wave interference. When N = 3, the wave vectors are separated by 120° and the incident-wave phases differ successively by 120°, producing a hexagonal-lattice pressure distribution referred to as an acoustic skyrmion interference field.[16,17]

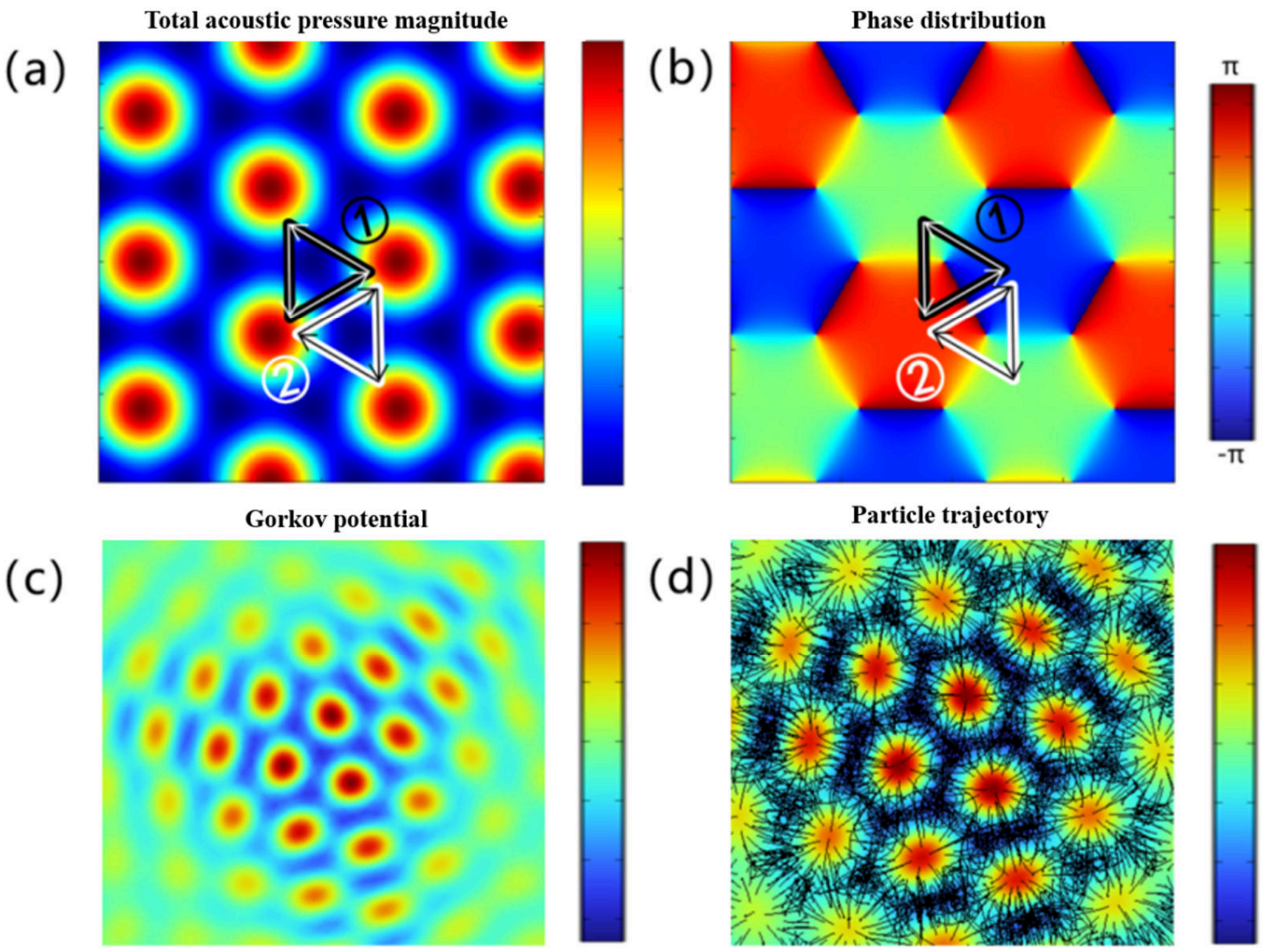

Fig. 3. Acoustic skyrmion interference field: (a) acoustic-pressure distribution; (b) phase distribution; (c) Gor'kov potential; (d) particle trajectories.

Within the acoustic skyrmion lattice, unstable high-intensity regions coexist with stable phase singularities, as shown in Fig. 3(a). At a phase singularity, the phase is undefined and the acoustic pressure is zero. Microparticles with a positive acoustic contrast factor are driven by the acoustic radiation force toward stable phase singularities. Fig. 3(b) further shows the spin texture of the lattice, with two counter-chiral spin states within each unit cell. Fig. 3(c) presents the Gor'kov potential in the acoustic skyrmion interference field. Its minima act as acoustic traps; the acoustic radiation force is obtained from the negative gradient of the potential, causing particles to accumulate at these traps.[19–21] The minima of the Gor'kov potential coincide with zero total acoustic pressure and the phase singularities. As illustrated by the trajectories in Fig. 3(d), initially dispersed particles assemble into a hexagonal lattice under acoustic actuation. Meanwhile, circulation of acoustic energy around each singularity generates tangential forces and torque, enabling combined orbital and spin motion.[22–24]

## 3. Structural design of the topological skyrmion-type acoustic-tweezer device

Based on the finite-element simulation results, a SAW platform was developed to verify microparticle manipulation. The device uses a Z-cut lithium niobate ($LiNbO_3$) crystal as the core substrate, which has strong piezoelectric coupling and favorable acoustic propagation characteristics. Its (001) plane has three symmetry axes, and the plane-wave intensity varies periodically with a 60° period, supporting comparable SAW amplitudes for multidirectional excitation. The SAW velocity on Z-cut $LiNbO_3$ is approximately 3990 m/s, and the relevant resonance band is concentrated near 10-12 MHz. Appropriate IDT design therefore enables efficient excitation and directional propagation of acoustic energy. After optimization, substrate anisotropy can also suppress unwanted scattering and improve field-control accuracy, providing a stable physical basis for multimode particle manipulation.[4,9,25,26]

The device employs a dual-ring, hierarchical IDT array. The outer ring comprises six low-frequency IDTs with a center frequency of 15.8 MHz and an angular spacing of 60°. Because the anisotropy of $LiNbO_3$ strongly suppresses SAWs away from preferred directions, the IDTs were aligned with the symmetry axes of the Z-cut substrate to maximize electromechanical coupling. Each IDT contains 50 finger pairs; both the finger width and the gap are 50 μm (consistent with the $\lambda/4$ design rule, where $\lambda$ is the acoustic wavelength), and the acoustic aperture is 10 mm. This geometry provides broad coverage and strong penetration for large-area particle patterning. The inner ring comprises eight vertically symmetric high-frequency IDTs with a center frequency of 38.8 MHz and an angular spacing of 30°. Each IDT contains 30 finger pairs; the electrode width and gap are optimized to 25 μm according to the wavelength, and the acoustic aperture is 3 mm. These high-frequency IDTs generate a high-resolution field for fine particle trapping and rotational manipulation.

The electrodes were fabricated by standard photolithography, evaporation, and lift-off. A Cr/Au bilayer was used: a 10 nm Cr adhesion layer and a 50 nm Au conductive layer. The total electrode thickness was maintained below 200 nm to avoid significant attenuation of SAW propagation.

A circular PDMS microfluidic chamber was positioned at the center of the Z-cut $LiNbO_3$ substrate. The chamber had an outer diameter of 2 mm, an inner diameter of 1.8 mm, and a height of 60 μm. It was molded from Sylgard 184 silicone elastomer and curing agent mixed at a 10:1 ratio. An ultrasonic couplant transferred SAW energy from the substrate into the chamber. One inlet and one outlet, each 100 μm in diameter, were incorporated for fluid injection and exchange.

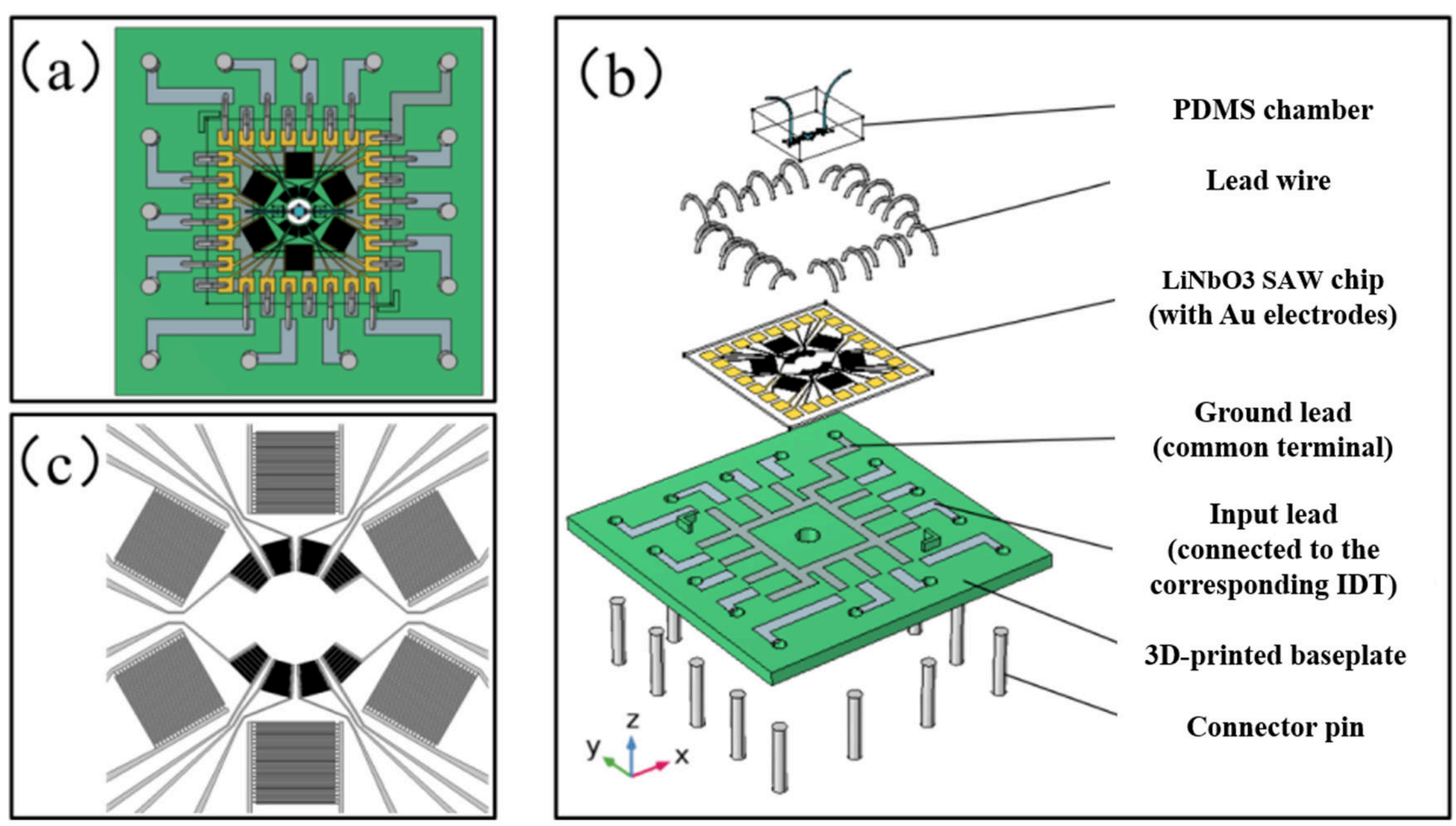

Fig. 4. Schematic of the topological skyrmion-type acoustic-tweezer device: (a) top view of the acoustofluidic chip; (b) assembly; (c) IDT electrode-layout mask.

As shown in Fig. 4(b), the disposable microfluidic chamber consists of three PDMS layers. The chamber height was approximately 60 µm, and the thin bottom sealing layer was approximately 30 µm thick. The top layer was fabricated by standard soft lithography and replica molding. For the bottom layer, a thin PDMS mixture was spin-coated onto a silicon wafer and baked at 65 °C for 20 min. The top layer was then placed over the bottom layer with the open chamber facing downward, followed by baking at 65 °C for 40 min to bond the layers and seal the chamber. The sealed chamber was peeled from the silicon wafer and cut to the required dimensions. Before each test, a drop of water was placed on the $LiNbO_3$ substrate as a couplant, after which the chamber was positioned directly over the target region. The water layer allowed SAW energy generated on the $LiNbO_3$ substrate to pass through the thin PDMS bottom layer into the microfluidic chamber. Because the chamber was not plasma-bonded to the substrate, it could be removed without damaging the $LiNbO_3$. The chamber was discarded after each test, allowing the substrate to be reused while preventing cross-contamination between experiments.

A 3D-printed resin baseplate (40 mm×40 mm×5 mm) supported the $LiNbO_3$ chip (15 mm×15 mm×0.5 mm). Pin interfaces and wire channels were reserved in the baseplate to ensure stable electrical connections.

Each IDT electrode was connected to an external signal generator through a gold-plated lead and assigned an independent signal channel for individual or combined excitation. A common ground port was provided to reduce electrical interference.

The fabricated acoustofluidic device and PDMS chamber are shown in Fig. 5.

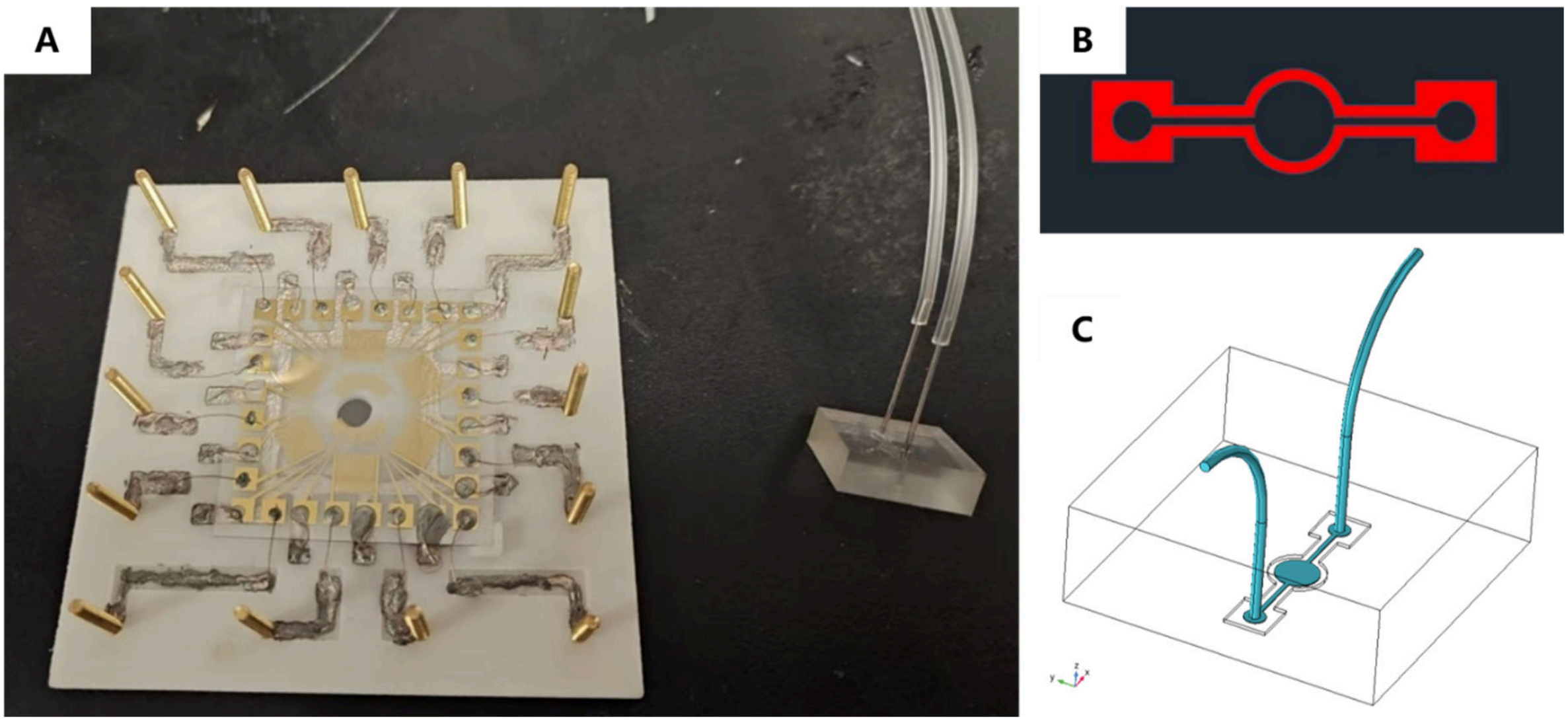

Fig. 5. The topological skyrmion-type acoustic-tweezer and PDMS chamber: (a) photograph of the acoustofluidic chip and PDMS chamber; (b) PDMS chamber mask; (c) PDMS chamber model.

The acoustic properties of the Z-cut $LiNbO_3$ substrate and the coordinated IDT-array design support multiple acoustic-field modes. Exciting a single IDT group generates a directional SAW that couples into the liquid to form a linear or vortex acoustic field. Combined excitation of multiple IDT groups uses wave interference to construct two-dimensional standing-wave fields or acoustic skyrmion fields. By switching the excited IDT combination and phase parameters, the wave-vector control afforded by the Z-cut substrate permits flexible field-mode conversion and multifunctional manipulation, including two-dimensional particle patterning, orbital rotation, and spin motion.[14,15,27,28]

## 4. Experimental Validation of Multifunctional Microparticle Manipulation Using Surface Acoustic Waves

### 4.1 Experimental preparation and platform setup

Monodisperse fluorescent polystyrene microspheres were purchased from Jiangsu Zhichuan Technology Co., Ltd. Two aqueous suspensions were used: 10 μm green-fluorescent microspheres and 5 μm red-fluorescent microspheres. Both were dispersed in deionized water at a solids concentration of 25 mg/mL, with 20 mL supplied per suspension. As shown in Fig. 6, the fabricated acoustic-tweezer chip was secured to the stage of an inverted fluorescence microscope (IX73, Olympus) using adhesive tape, and a disposable microfluidic chamber was positioned over the lithium niobate piezoelectric substrate. The microparticle suspension was introduced into the chamber using a disposable syringe. The SAW excitation signal (15.8 MHz and 10 Vpp) was generated by a dual-channel arbitrary waveform generator (DG1022U, RIGOL), amplified with a gain of 20 dB using a power amplifier (ATA-1200B, AIGTEK), and applied to the interdigital transducers (IDTs). Individual IDTs were activated by connecting the signal line to the corresponding pins, enabling rapid and precise reconfiguration of the SAW field and thereby controlling particles or cells within the microfluidic chamber. Images and videos were acquired using the inverted fluorescence microscope and cellSens Standard imaging software (Olympus). The experimental workflow is illustrated in Fig. 6.

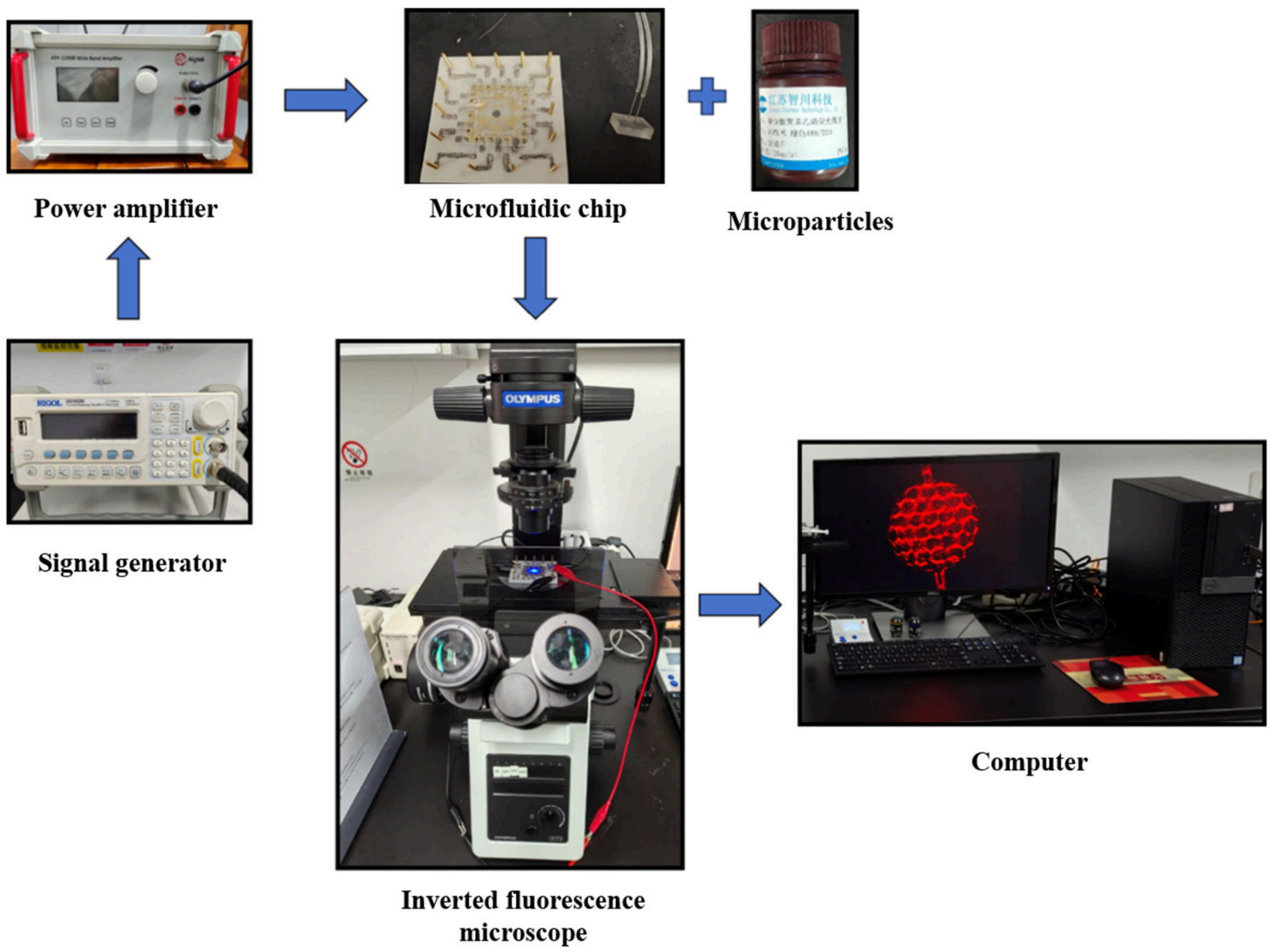

Fig. 6. Workflow of the microfluidic experimental platform.

### 4.2 Experimental results and analysis

As shown in Fig. 7, particles initially arranged in one-dimensional lines under a sustained standing-wave field were manipulated by switching the phase difference between the two incident waves from $\Delta\varphi = 0$ instantaneously to $\Delta\varphi = \pi$. This switch caused the 10 μm fluorescent microparticles at the original nodes to converge; reversing the switch caused them to disperse.

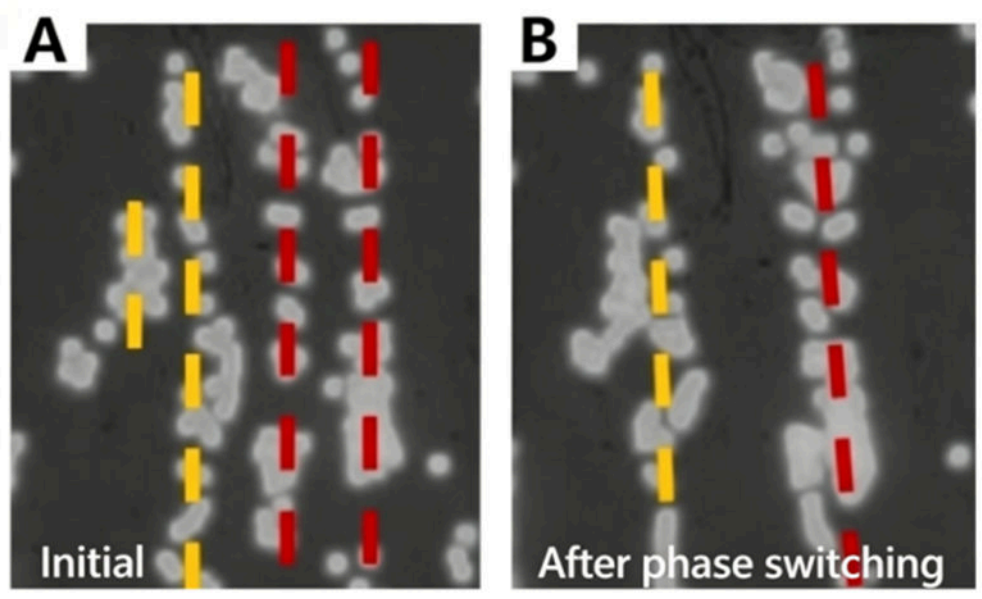

Fig. 7. Phase-switching-induced convergence and separation of microparticles: (a) fluorescent microparticles in the initially dispersed state; (b) convergence after phase switching.

As shown in Fig. 8, when SAWs were excited at 15.8 MHz by two pairs of peripheral interdigital transducers (IDTs) oriented at an angle of 60° to each other, a two-dimensional standing-wave field was generated in the central region. The acoustic energy was coupled into the microfluidic chamber through an ultrasonic coupling medium, producing a periodic two-dimensional acoustic-pressure distribution within the liquid. Under the resulting acoustic radiation force, 10 μm polystyrene microspheres suspended in the liquid were driven toward the equilibrium positions and assembled into a two-dimensional particle array.

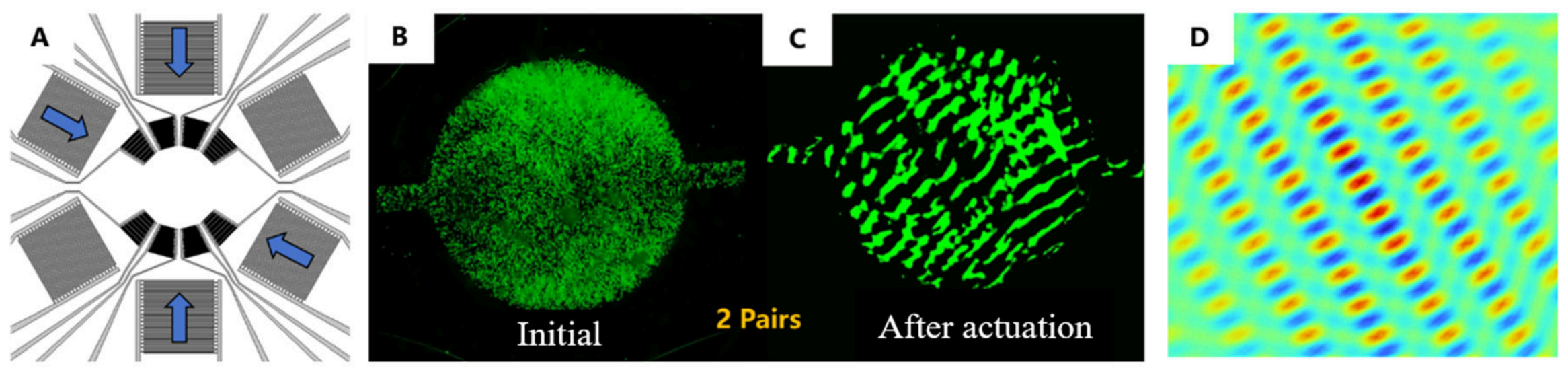

Fig. 8. Microparticle patterning in a two-dimensional standing-wave field generated by two IDT pairs separated by 60°: (a) excited IDTs; (b) initial microchannel state with uniformly dispersed 10 μm green-fluorescent microparticles; (c) two-dimensional particle pattern after acoustic actuation; (d) Gor'kov potential.

As shown in Fig. 9, three pairs of peripheral interdigital transducers (IDTs), arranged at angular intervals of 60°, were driven at 15.8 MHz to excite surface acoustic waves (SAWs). The phases applied to two pairs of IDTs were set to 0, whereas that applied to the remaining pair was set to π. The resulting interference generated a two-dimensional standing-wave field in the central region. The acoustic energy was coupled into the microfluidic chamber through an ultrasonic coupling medium, producing a distinct periodic two-dimensional acoustic-pressure distribution within the liquid. Under the resulting acoustic radiation force, the suspended 10 μm polystyrene microspheres migrated toward stable equilibrium positions and assembled into an interwoven two-dimensional lattice pattern.

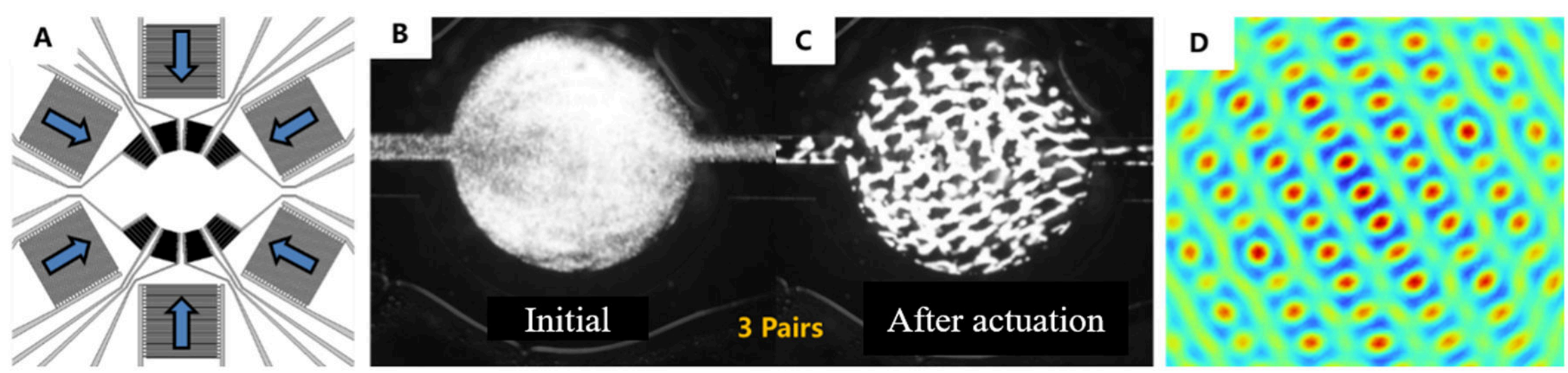

Fig. 9. Microparticle patterning in a two-dimensional standing-wave field generated by three IDT pairs separated by 60°: (a) excited IDTs; (b) initial microchannel state with uniformly dispersed 10 μm fluorescent microparticles; (c) two-dimensional woven-grid particle pattern after acoustic actuation; (d) Gor'kov potential.

The relatively weak contrast of the observed two-dimensional particle pattern is attributed primarily to coupling instability. Heating caused by SAW excitation gradually evaporated the water used as the couplant. The PDMS chamber then relied on direct mechanical contact with the piezoelectric substrate to receive acoustic energy from the IDTs. Because this simple contact was not mechanically robust, external disturbances could partially separate the interfaces or produce poor local contact. Consequently, only limited acoustic energy entered some regions of the chamber, reducing the local field strength. The corresponding acoustic radiation force was therefore too weak to displace the microparticles rapidly and effectively.

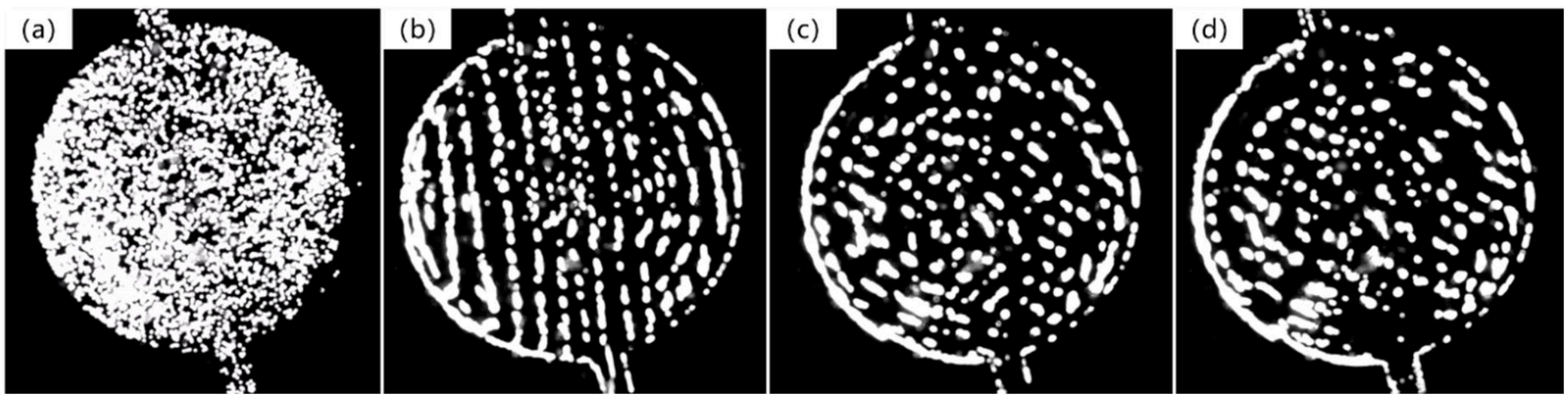

Fig. 10. Sequential adjustment of the standing-wave field: (a) initially dispersed microparticles; (b) one IDT pair activated; (c) two IDT pairs activated; (d) three IDT pairs activated.

The particle concentration in the microfluidic chamber was subsequently reduced, as shown in Fig. 10(a). The particles were initially dispersed. Activating one IDT pair produced a one-dimensional line pattern, shown in Fig. 10(b); activating two IDT pairs produced the pattern shown in Fig. 10(c); activating three pairs produced the pattern shown in Fig. 10(d). These results demonstrate dynamic, simultaneous control of a large number of particles by adjusting the activated IDTs.

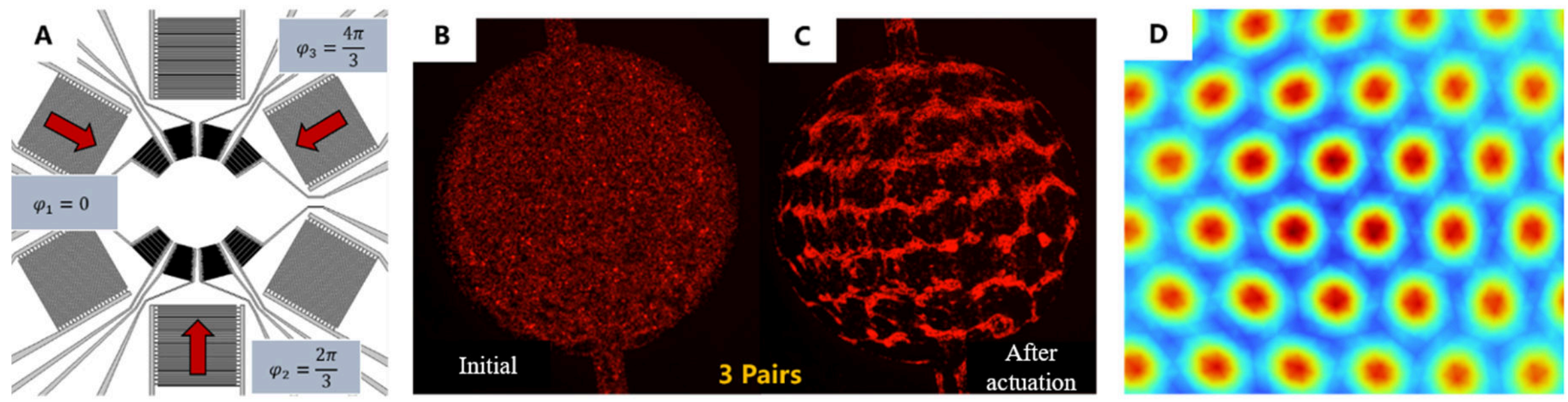

Fig. 11. Microparticle patterning in a two-dimensional acoustic skyrmion interference field generated by three IDTs separated by 120° with successive phase differences of 120°: (a) excited IDTs and applied phases; (b) initial microchannel state with uniformly dispersed 5 µm red-fluorescent microparticles; (c) two-dimensional hexagonal particle lattice after acoustic actuation; (d) Gor'kov potential.

As shown in Fig. 11, three outer-ring IDTs operating at 15.8 MHz and separated by 120° were used to excite SAWs. The initial phases of the first, second, and third incident waves were 0°, 120°, and 240°, respectively. The resulting SAWs formed a two-dimensional acoustic skyrmion interference field in the central region.[16,17] Acoustic energy entered the microfluidic chamber through the ultrasonic couplant and established the corresponding periodic pressure distribution in the liquid. Under the acoustic radiation force, 5 µm polystyrene microspheres assembled into a two-dimensional hexagonal lattice.

Although a conventional simple IDT structure can also produce two-dimensional particle arrays, the independently addressable, dual-ring hierarchical IDT array developed here offers three principal advantages. (1) Multifunctionality without chip replacement: conventional systems often require IDT chips of different geometries to switch from particle patterning to rotation, whereas the present device changes functions within seconds by electronically selecting different combinations. The outer low-frequency IDTs provide large-area patterning, while the inner high-frequency IDTs support fine rotational manipulation. (2) Generation of topological acoustic fields: simple IDTs generally cannot produce acoustic skyrmions or related fields carrying angular momentum, which provide mechanisms for particle spin and orbital motion.[16,18,22–24] (3) Robustness and reconfigurability: topologically structured multiwave fields can resist localized perturbations, while phase modulation dynamically reconfigures the particle-control landscape.[14–17,28] The added device complexity is therefore the necessary cost of functional diversity, flexibility, and robustness.

Because of current equipment limitations, the experiments used a single type of suspension (fluorescent microparticles in water) to verify particle patterning in two-dimensional standing-wave fields, phase-controlled convergence and separation, and manipulation in the acoustic skyrmion interference field.

## 5. Conclusions

To address the susceptibility to disturbances and insufficient trapping stability of conventional SAW microparticle-manipulation fields, this work introduces topological acoustics into a SAW system and develops a topological acoustic skyrmion tweezer based on a Z-cut $LiNbO_3$ substrate. A systematic study encompassing theoretical analysis, device design, and functional verification was completed. The principal conclusions are as follows:

1. A theoretical model of the topological acoustic skyrmion field was established from the principle of multiwave interference. The total pressure field produced by coherent superposition of multiple plane waves was derived, revealing how three waves separated by 120° in wave-vector direction and by successive 120-degree phase offsets form a topological acoustic skyrmion lattice. The correspondence among phase singularities, spin textures, and the Gor'kov potential was clarified. Controlling the wave-vector directions and incident phase differences enables construction of a field with topological protection, a prescribed energy distribution, and defined spin characteristics, thereby providing the theoretical basis for stable particle trapping and controlled rotation.

2. A dual-ring, hierarchical IDT array was designed to provide the structural basis for precise formation of the topological acoustic field. Six low-frequency outer-ring IDTs (15.8 MHz) generate coherent SAW interference over a large area and support formation of an extended acoustic skyrmion lattice. Eight high-frequency inner-ring IDTs (38.8 MHz) provide a higher-resolution acoustic field for fine particle trapping and rotational manipulation. The device integrates an annular PDMS microfluidic chamber and a 3D-printed resin baseplate. Independent signal channels provide separate amplitude and phase control for each IDT, ensuring flexibility in field construction and stable electrical connections.

3. Experiments verified microparticle manipulation by the topological acoustic skyrmion field. Selective excitation of the appropriate IDTs with matched phase parameters produced a stable acoustic skyrmion interference field in the microfluidic chamber. Potential wells at the phase singularities enabled site-specific trapping and regular hexagonal-lattice assembly. The intrinsic spin texture of the field exerted acoustic radiation torque on the particles, allowing controlled spin and orbital motion. Switching among IDT excitation combinations also generated two-dimensional standing-wave fields of different geometries and produced point-array particle patterns, confirming the field-reconfiguration capability of the device.

This study integrates topological acoustic structures with SAW micromanipulation and translates the stability and robustness associated with topological structuring into practical microparticle-control functions. The proposed approach provides an alternative to conventional interference fields that are sensitive to disturbances. Owing to the favorable acoustic properties of the Z-cut $LiNbO_3$ substrate, the device combines high spatial resolution with reconfigurable field control and can accommodate microparticles of different sizes. It therefore provides a new route to high-precision manipulation in microfluidic chips, with potential applications in single-cell control, three-dimensional cell assembly, high-throughput screening, virus isolation, and micro/nanodevice assembly.[10–13,24]